\documentclass[twocolumn,english,aps,pra,superscriptaddress,showpacs,tightenlines]{revtex4}
\usepackage[T1]{fontenc}
\usepackage[latin9]{inputenc}
\usepackage{amsmath}
\usepackage{graphicx}
\usepackage{amssymb}
\usepackage{CJK}

\input{tcilatex}
\begin{document}


\title{Adiabatic Creation of Atomic Squeezing in Dark States vs. Decoherences }

\author{Z. R. Gong)}

\affiliation{Institute of Theoretical Physics, Chinese Academy of Sciences, Beijing
100190, China}

\author{Xiaoguang Wang}

\affiliation{Zhejiang Institute of Modern Physics, Department of Physics, Zhejiang
University, Hangzhou 310027, China}

\author{C. P. Sun}

\affiliation{Institute of Theoretical Physics, Chinese Academy of Sciences, Beijing
100190, China}

\email{suncp@itp.ac.cn}
\begin{abstract}
We study the multipartite correlations of the multi-atom dark
states, which are characterized by the atomic squeezing beyond the
pairwise entanglement. It is shown that, in the photon storage
process with atomic ensemble via electromagnetically induced
transparency (EIT) mechanism, the atomic squeezing and the pairwise
entanglement can be created by adiabatically manipulating the Rabi
frequency of the classical light field on the atomic ensemble. We
also consider the sudden death for the atomic squeezing and the
pairwise entanglement under various decoherence channels. An optimal
time for generating the greatest atomic squeezing and pairwise
entanglement is obtained by studying in details the competition
between the adiabatic creation of quantum correlation in the atomic
ensemble and the decoherence that we describe with three typical
decoherence channels.
\end{abstract}


\maketitle

\section{\label{sec:one}Introduction}

Atomic ensemble can serve as a quantum memory for storing the quantum
information of photons~\cite{Hau,Lukin-1,Lukin-2,Lukin-3} where the memory
elements can be described as the quasi-spin wave excitations in atomic
ensemble~\cite{2003}, or Dicke type collective states~\cite{Dicke}. Recent
experiments have demonstrated that such storable and quantum memory could
have long lifetime for the use in long-distance quantum communication~\cite%
{Pan}. When using the magnetically insensitive clock transition in atomic
rubidium confined in a one-dimensional optical lattice, quantum memory
lifetime can exceed 3 ms~\cite{Kuzmich}.

The relevant quantum storage process could be implemented through the
many-particle enhancement of the absorption cross section with the
adiabatic-passage techniques~\cite{APT}. Through the electromagnetically
induced transparency (EIT) mechanism, the adiabatic manipulation utilizes
the dark polariton $|d_{n}(\theta)\rangle$ (a photon-dressed collective
atomic state). With adiabatically changing parameter $\theta\in[0,\pi/2]$,
the quantum state of photon $|P\rangle=\sum_{n} c_{n}|n\rangle$ (a
superposition of photon Fock states $|n\rangle$) can be adiabatically
converted to a state of the collective atom excitation $|M\rangle=$ $%
\sum_{n} c_{n}|M_{n}\rangle$, which is a superposition of the collective
atomic states $|M_{n}\rangle$ with the same coefficients $c_{n}$ as that in
the photon state $|P\rangle$. Mathematically, this coherent conversion of
the photon state is an associated mapping
\begin{equation}
|P\rangle\otimes|d^{\prime}\rangle=\sum c_{n}|d_{n}(0)\rangle\rightarrow\sum
c_{n}|d_{n}(\frac{\pi}{2})\rangle=|p^{\prime}\rangle\otimes|M\rangle,
\label{eq:1-1}
\end{equation}
where $|d^{\prime}\rangle$ and $|p^{\prime}\rangle$ are the initial state of
the memory and the final state of photon respectively. Obviously, here only
used are the macroscopically coherent properties of each collective
excitation, which is described by individual collective state $%
|M_{n}\rangle, $ rather than the internal entanglement and quantum
correlation \ in the single collective state $|M_{n}\rangle$ as well as $%
|d_{n}(\theta)\rangle$.

In this paper we will pay attention to the latter by considering the
multipartite correlation measured by spin squeezing in $|M_{n}\rangle$,
which is obtained from $|d_{n}(\theta)\rangle$ by adiabatic manipulation for
the dark state $|d_{n}(\theta)\rangle$ as
\begin{equation}
|n\rangle\otimes|0\rangle=|d_{n}(0)\rangle\rightarrow|d_{n}(\frac{\pi}{2}%
)\rangle=|0\rangle\otimes|M_{n}\rangle.  \label{eq:1-2}
\end{equation}
We study in details the dynamic competition between the adiabatic creation
of quantum correlation in the atomic ensemble and the decoherence that we
describe with three typical decoherence channels. There are two time scales $%
t_{1}$ and $t_{2}$ depicting such competition, where $t_{1}$ represents the
adiabatic time limited by the adiabatic conditions and $t_{2}$ represents
the decoherence time. Actually, if we regard the atomic ensemble as a
macroscopic object in large N limit, quantum decoherence scenario seems to
result in the quantum disappearance due to noisy ambient environments~\cite%
{Noisy} or the internal random motions~\cite{Random1,Random2}. A surprising
discovery about the pairwise entanglement is that it suddenly die due to
influence of environment~\cite{Yuting,Wang-1}. In this paper, we also
consider this sudden death phenomena of the pairwise entanglement as well as
the atomic squeezing for quantum correlation for more than two particles. We
compare the time scales of various sudden death processes with the speed of
adiabatic manipulations for creating such squeezing as well as the pairwise
entanglement.

Actually, creating spin squeezing in a single Dicke like state is the
crucial step for the precision measurements based on many-atom spectroscopy.
With such controlled production of complex entangled states of matter and
light, squeezing the fluctuations via entanglement between 2-level atoms can
improve the precision of sensing, clocks, metrology, and spectroscopy~\cite%
{Polzik}. This paper seems to provide a simple scheme for adiabatic creation
of atomic squeezing. This adiabatic passage manipulation seems feasible in
creating the spin squeezing, but it requires an ideal technique of single
photon source, which can initially produce the single Fock states $%
|n\rangle. $ Therefore, our present theoretical proposal may meet the
difficulty for physical implementation based on the existing technology.

This paper is arranged as follows. In Sec.~\ref{sec:two}, we study
the multipartite correlations characterized by the atomic squeezing
beyond the pairwise entanglement in the dark states. In
Sec.~\ref{sec:two}, we study the relationship between the
concurrence and the spin squeezing under adiabatic manipulation, and
indicate that the photon statistics varies in the same way of the
spin squeezing. In Sec.~\ref{sec:four}, the sudden death of the
atomic squeezing and the pairwise entanglement are considered by
introducing three typical decoherence channels. The optimal times
for generating the atomic squeezing and the pairwise entanglement
are obtained in Sec.~\ref{sec:five}. We conclude in
Sec.~\ref{sec:five}.

\section{\label{sec:two}Adiabatically creating squeezing in dark states}

%
\begin{figure}[ptb]
\begin{centering}
\includegraphics[bb=100 361 494 770,clip,width=3in]{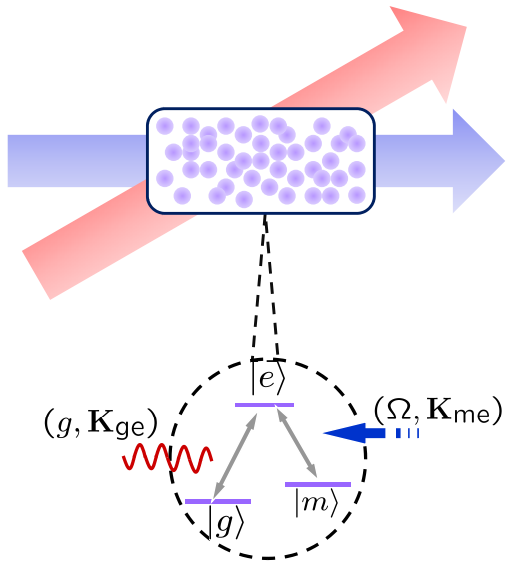}
\par\end{centering}
\caption{(Color online) Schematic illustration of the atomic ensemble
interacting with two light fields: the classical light field (denoted by the
blue arrow) with Rabi frequency $\Omega$ and the wave vector $K_{\mathrm{me}%
} $ and the quantum light field (denoted by the red arrow) with the coupling
constant $g$ and the wave vector $K_{\mathrm{ge}}$. Each $\Lambda-$type atom
confined in the rectangular container has the same excited state $%
\left|e\right\rangle $, the relative ground state $\left|g\right\rangle $
and the metastable state $\left|m\right\rangle $.}
\label{fig:fig1}
\end{figure}


\subsection{Dark states}

As shown in Fig.~\ref{fig:fig1}, the model we consider consists of atomic
ensemble with $N$ $\Lambda -$type atoms, which have the excited state $%
\left\vert e\right\rangle $, the relative ground state $\left\vert
g\right\rangle $ and the metastable state $\left\vert m\right\rangle $. For
convenience we assume all the atoms have the same energy spacings for the
above three atomic states. For the cold atomic ensemble, the Doppler width
broadening of the atoms can be depressed effectively. These atoms interact
with two single-mode light fields: one is quantized radiation mode (with
coupling constant $g$ and annihilation operator $a$) to lead the
approximately resonant transition from $\left\vert e\right\rangle $ to $%
\left\vert g\right\rangle $; the other is an exact classical field with Rabi
frequency $\Omega $ to lead the transition from $\left\vert e\right\rangle $
to $\left\vert m\right\rangle $. The quantum dynamics of the total system is
described by the following Hamiltonian in the interaction picture~\cite{2003}%
\begin{align}
H& =ga\sum_{\mathbf{j}=1}^{N}\exp \left( i\mathrm{\mathbf{K}}_{\mathrm{ge}%
}\cdot \mathbf{j}\right) \sigma _{\mathrm{ge}}^{\mathbf{j}}+  \notag \\
& \Omega \sum_{\mathbf{j}=1}^{N}\exp \left( i\mathrm{\mathbf{K}}_{\mathrm{me}%
}\cdot \mathbf{j}\right) \sigma _{\mathrm{me}}^{\mathbf{j}}+h.c.,
\label{eq:2-1-1}
\end{align}%
where $\mathrm{\mathbf{K}}_{\mathrm{ge}}$ and $\mathrm{\mathbf{K}}_{\mathrm{%
me}}$ are, respectively, the wave vectors of the quantum and the classical
light fields. We have introduced the quasi-spin operators $\sigma _{\alpha
\beta }^{\mathbf{j}}\equiv \left\vert \alpha \right\rangle _{\mathbf{jj}%
}\left\langle \beta \right\vert (\alpha ,\beta =e,g,m)$ for $\alpha \neq
\beta $ in above Hamiltonian to describe the transitions among the levels of
$\left\vert e\right\rangle ,$ $\left\vert g\right\rangle $ and $\left\vert
m\right\rangle .$

For EIT, to find the degenerate class of the eigenstates $%
\left|d_{n}\left(\theta\right)\right\rangle $ satisfying
\begin{equation}
H\left|d_{n}\left(\theta\right)\right\rangle =0,  \label{eq:2-1-2}
\end{equation}
we introduce the dark state polariton operator%
\begin{equation}
D^{\dagger}\left(\theta\right)=a^{\dagger}\cos\theta-C^{\dagger}\sin\theta,
\label{eq:2-1-3}
\end{equation}
which obviously mix the electromagnetic field with creation operator $%
a^{\dagger}$ and the quasi-spin wave with corresponding operator
\begin{equation}
C^{\dagger}=\frac{1}{\sqrt{N}}\sum_{\mathbf{j}=1}^{N}\exp\left(i\mathbf{K}%
\cdot\mathbf{j}\right)\sigma_{mg}^{\mathbf{j}}.  \label{eq:2-1-4}
\end{equation}
The angle $\tan\theta=g\sqrt{n}/\Omega$ can be adiabatically manipulated in
the quantum storage process, and $\mathbf{K=\mathrm{\mathbf{K}}_{\mathrm{ge}}%
}-\mathrm{\mathbf{K}}_{\mathrm{me}}$ is the wave vector difference between
the quantum and the classical fields. We construct the degenerate class of
the eigenstates $\left|d_{n}\left(\theta\right)\right\rangle $ as

\begin{equation}
\left|d_{n}\left(\theta\right)\right\rangle =\frac{A(n,N,\theta)}{\sqrt{n!}}%
\left[D^{\dagger}\left(\theta\right)\right]^{n}\left|\mathbf{0}\right\rangle
,  \label{eq:2-1-5}
\end{equation}
where the normalization constant%
\begin{align}
A(n,N,\theta) & =\sqrt{\frac{\left(N-n\right)!N^{n}}{N!\sin^{2n}2\theta}}%
\times  \notag \\
& \left[\text{ }_{1}F_{1}(-n;N-n+1;-N\cot^{2}\theta)\right]^{-\frac{1}{2}}
\label{eq:2-1-6}
\end{align}
reduces to $1$ when the atom number $N$ is much larger than $n$. Here, $n$
is the excitation number of the atomic ensemble, the vacuum state
\begin{equation}
\left|\mathbf{0}\right\rangle =\left|0\right\rangle \otimes\left|\mathbf{%
\downarrow}\right\rangle  \label{eq:2-1-7}
\end{equation}
is the direct product of the photonic vacuum state $\left|0\right\rangle $
and the quasi-spin ground state $\left|\mathbf{\downarrow}\right\rangle
\equiv\prod_{\mathbf{j=1}}^{N}\left|g\right\rangle _{\mathbf{j}}.$ Here,$%
\text{ }_{1}F_{1}(a;b;c)$ is the Kummer hypergeometric function~\cite{Kummer}%
. Eq. (\ref{eq:2-1-2}) indicates that $\left|d_{n}\left(\theta\right)\right%
\rangle $ is totally cancelled by the interaction Hamiltonian, and thus is
called a dark state or a dark state polariton.

Obviously, the dark state $\left|d_{n}\left(\theta\right)\right\rangle $ is
photon Fock state $\left|n\right\rangle $ when $\theta=0$ and quasi-spin
wave state when $\theta=\pi/2$. In the low excitation limit as $n\ll N$, $%
C^{\dagger}$ is an approximately bosonic operator as well as $%
D^{\dagger}\left(\theta\right)$.

With the help of the dark states, by adiabatically manipulating the
classical field the quantum information of the photon with quantum state $%
\left|P\right\rangle =\sum_{n}c_{n}\left|n\right\rangle $ can be coherently
converted into the atomic ensemble as $\left|M\right\rangle
=\sum_{n}c_{n}\left|M_{n}\right\rangle $, where $\left|n\right\rangle $ is
the Fock state of photon and
\begin{align}
\left|M_{n}\right\rangle & =\left|d_{n}\left(\theta=\frac{\pi}{2}%
\right)\right\rangle  \notag \\
& =\left(-1\right)^{n}\sqrt{\frac{N^{n}(N-n)!}{N!n!}}\left[C^{\dagger}\right]%
^{n}\left|\mathbf{0}\right\rangle  \label{eq:2-1-8}
\end{align}
is the collective atomic state. This coherent conversion is an associated
mapping
\begin{equation}
|P\rangle\otimes|\downarrow\rangle=\sum c_{n}|d_{n}(0)\rangle\rightarrow\sum
c_{n}|d_{n}(\frac{\pi}{2})\rangle=|0\rangle\otimes|M\rangle,
\label{eq:2-1-9}
\end{equation}
which only use the macroscopically coherent properties of each collective
excitation described by individual collective state $\left|M_{n}\right%
\rangle .$ We notice that $\left|M_{n}\right\rangle $ is actual multi-atom
state with multipartite quantum correlations. For two-particle case, we call
it internal entanglement.

The internal entanglement and the quantum correlation in the single
collective state $\left|M_{n}\right\rangle $ as well as $|d_{n}(\theta)%
\rangle$ also play important role in quantum storage process. Now we
consider the quantum correlation measured by spin squeezing of those
quasi-spins in $\left|M_{n}\right\rangle $, which is obtained from $%
|d_{n}(\theta)\rangle$ by adiabatic manipulation%
\begin{equation}
|n\rangle\otimes|0\rangle=|d_{n}(0)\rangle\rightarrow|d_{n}(\frac{\pi}{2}%
)\rangle=|0\rangle\otimes|M_{n}\rangle.  \label{eq:2-1-10}
\end{equation}

\subsection{Atomic squeezing in dark states without decoherence}

The above dark state mixes the electromagnetic field and the quasi-spins
defined by the two levels $\left|g\right\rangle $ and $\left|m\right\rangle $
of the atomic ensemble. The collective operators
\begin{equation}
J_{\alpha}=\frac{1}{2}\sum_{\mathbf{j}=1}^{N}\sigma_{\alpha}^{\mathbf{j}%
},(\alpha=x,y,z)  \label{eq:2-2-1}
\end{equation}
can be generally used to demonstrate the exchange symmetry and dynamic
properties of the quasi-spin system. To calculate the averages of the
collective operators, the reduced density matrix of the quasi-spins is
obtained by tracing over the degree of freedom of the photons as $\rho_{%
\mathrm{a}}^{\mathrm{r}}=\mathrm{Tr}_{\mathrm{p}}\left\vert
d_{n}\left(\theta\right)\right\rangle \left\langle
d_{n}\left(\theta\right)\right\vert $ or%
\begin{align}
\rho_{\mathrm{a}}^{\mathrm{r}} & =\sum_{k=0}^{n}\frac{n!\left(\cos\theta%
\right)^{2k}\left(\sin\theta\right)^{2n-2k}}{k!(n-k)!}\frac{%
\left|A(n,N,\theta)\right|^{2}N!}{N^{k}(N-n+k)!}\times  \notag \\
& U\left\vert n-k-\frac{N}{2}\right\rangle \left\langle n-k-\frac{N}{2}%
\right\vert U^{\dagger},  \label{eq:2-2-2}
\end{align}
where
\begin{equation}
\left\vert k-\frac{N}{2}\right\rangle \equiv\sqrt{\frac{\left(N-k\right)!}{%
N!k!}}\left(J_{+}\right)^{k}\left\vert \downarrow\right\rangle
\label{eq:2-2-3}
\end{equation}
is a symmetric Dicke state and the collective operator
\begin{equation}
J_{+}=\sqrt{N}U^{\dagger}C^{\dagger}U  \label{eq:2-2-4}
\end{equation}
is a unitary transformation of the quai-spin wave operator with the unitary
matrix%
\begin{equation}
U=\exp\left[-i\sum_{j=1}^{N}\mathbf{K\cdot j}\frac{\left(\sigma_{z}^{\mathbf{%
j}}+1\right)}{2}\right].  \label{eq:2-2-5}
\end{equation}

Since only is the $z-$component quasi-spin operator contained in the unitary
transformation, which will never change the quantum number of the symmetric
Dicke bases, the density matrix in Eq. (\ref{eq:2-2-2}) only has the
diagonal elements. Thus the averages of $J_{x}$ and $J_{y}$ vanish as $%
\left\langle J_{x}\right\rangle =\left\langle J_{y}\right\rangle =0.$ In
this situation, only the $z-$component collective operator $J_{z}$ survives
as
\begin{align}
\left\langle J_{z}\right\rangle & =\left.\left\langle J_{z}\right\rangle
\right|_{\theta=\frac{\pi}{2}}+\delta J_{z},  \label{eq:2-2-6}
\end{align}
where $\left.\left\langle J_{z}\right\rangle \right|_{\theta=\pi/2}=n-N/2$
is mean $z-$component spin of the Dicke state $\left\vert n-N/2\right\rangle
$ and
\begin{equation}
\delta J_{z}=-\frac{nN\cot^{2}\theta_{1}F_{1}(1-n;N-n+2;-N\cot^{2}\theta)}{%
(N-n+1)_{1}F_{1}(-n;N-n+1;-N\cot^{2}\theta)}  \label{eq:2-2-7}
\end{equation}
is the deviation of the $z-$component spin. The fact that $\left\langle
J_{z}\right\rangle $ does not vanish means that the mean spin is along the $%
z-$direction. Since $\delta J_{z}$ vanishes when $\theta=\pi/2$, its value
can measure the mixture of the electromagnetic field and the quasi-spin wave.

The spin squeezing has several measures, and we only list three typical and
related ones as follows:\cite{KU,Wineland,Toth,Wang-1}

\begin{subequations}
\begin{align}
\xi_{1}^{2} & =\frac{4\left(\Delta J_{\perp}\right)_{\min}^{2}}{N},
\label{eq:2-2-8-a} \\
\xi_{2}^{2} & =\frac{N^{2}}{4\left\langle \mathbf{J}^{2}\right\rangle }%
\xi_{1}^{2},  \label{eq:2-2-8-b} \\
\xi_{3}^{2} & =\frac{\lambda_{\min}}{\left\langle \mathbf{J}%
^{2}\right\rangle -\frac{N}{2}}.  \label{eq:2-2-8-c}
\end{align}
In Eq.(\ref{eq:2-2-8-a}), the minimization is over all the direction denoted
by $\perp$, which is perpendicular to the mean spin direction $\left\langle
\mathbf{J}\right\rangle /\left\langle \mathbf{J}^{2}\right\rangle .$ For the
quasi-spins in the dark state, the $\xi_{1}^{2}$ and $\xi_{2}^{2}$ actually
measure the spin squeezing of the atomic ensemble in the $x-y$ plane. In Eq.(%
\ref{eq:2-2-8-c}), $\lambda_{\min}$ is the minimal eigenvalue of matrix
\end{subequations}
\begin{equation}
\Gamma=(N-1)\gamma+\mathbf{C},  \label{eq:2-2-9}
\end{equation}
where
\begin{equation}
\gamma_{kl}=\mathbf{C}_{kl}-\langle J_{k}\rangle\langle J_{l}\rangle\text{ }%
(k,l\in\{x,y,z\})  \label{eq:2-2-10}
\end{equation}
is the covariance matrix and
\begin{equation}
\mathbf{C}_{kl}=\frac{1}{2}\langle J_{l}J_{k}+J_{k}J_{l}\rangle
\label{eq:2-2-11}
\end{equation}
is the global correlation matrix. Parameters $\xi_{1}^{2},\xi_{2}^{2},$ and $%
\xi_{3}^{2}$ were defined by Kitagawa and Ueda, Wineland et al., and Toth et
al., respectively. If $\xi_{2}^{2}<1$ $(\xi_{3}^{2}<1),$ the spin squeezing
occurs, and we can safely say that the multipartite state is entangled \cite%
{Toth,Sorensen}. Although we cannot say that the squeezed state according to
$\xi_{1}^{2}<1$ is entangled, it is indeed closely related to quantum
entanglement \cite{Wang-2}.

For the quasi-spins in the dark state, since the averages $\left\langle
J_{x}J_{z}\right\rangle =\left\langle J_{y}J_{z}\right\rangle =0$ vanish as
the same reason of the vanishing of $\left\langle J_{x}\right\rangle $ and $%
\left\langle J_{y}\right\rangle $, the $\xi_{1}^{2}$ and $\xi_{3}^{2}$ are
simplified as~\cite{Wang-1,Wang-2}
\begin{subequations}
\begin{align}
\xi_{1}^{2} & =\frac{2}{N}\left(\left\langle
J_{x}^{2}+J_{y}^{2}\right\rangle -\left\vert \left\langle
J_{-}^{2}\right\rangle \right\vert \right),  \label{eq:2-2-12-a} \\
\xi_{3}^{2} & =\frac{\min\{\xi_{1}^{2},\varsigma^{2}\}}{\frac{4}{N^{2}}%
\left\langle \mathbf{J}^{2}\right\rangle -\frac{2}{N}},  \label{eq:2-2-12-b}
\end{align}
where
\end{subequations}
\begin{equation}
\varsigma^{2}=\frac{4}{N^{2}}\left[N\left(\Delta
J_{z}\right)^{2}+\left\langle J_{z}\right\rangle ^{2}\right]
\label{eq:2-2-13}
\end{equation}
characterizes the spin squeezing along the $z-$direction. Because $%
\left\langle J_{-}^{2}\right\rangle =0$ for the quasi-spins in the dark
state, $\xi_{1}^{2}$ and $\xi_{3}^{2}$ are always greater than $1$ and thus
there is no spin squeezing of the atomic ensemble in the $x-y$ plane. In the
following discussions, we only use
\begin{align}
\xi_{3}^{2} & =\frac{\varsigma^{2}}{\frac{4}{N^{2}}\left\langle \mathbf{J}%
^{2}\right\rangle -\frac{2}{N}}  \label{eq:2-2-14}
\end{align}
to characterize the $z-$component spin squeezing of the atomic ensemble in
the dark states.

We consider the special case that when the wave vector difference $\mathbf{K}
$ between the quantum and the classical field is zero. Since $%
\left.U\right|_{\mathbf{K}=0}$ is identity matrix, the quai-spin wave
operator in Eq. (\ref{eq:2-2-4}) actually is proportional to the collective
operator and the reduced density matrix in Eq. (\ref{eq:2-2-2}) is spanned
by the symmetric Dicke states with total quasi-spin $\mathbf{J}=N/2$. Thus
\begin{equation}
\left.\left\langle \mathbf{J}^{2}\right\rangle \right|_{\mathbf{K}=0}=\frac{N%
}{2}\left(\frac{N}{2}+1\right),  \label{eq:2-2-15}
\end{equation}
and the straightforward calculation gives
\begin{subequations}
\begin{align}
\left.\xi_{3}^{2}\right|_{\mathbf{K}=0} & =\varsigma^{2}=\frac{4}{N^{2}}%
\left[N\left\langle J_{z}^{2}\right\rangle +\left(1-N\right)\left\langle
J_{z}\right\rangle ^{2}\right],  \label{eq:2-2-16-a} \\
\left\langle J_{z}^{2}\right\rangle & =\left.\left\langle
J_{z}^{2}\right\rangle \right|_{\theta=\frac{\pi}{2}}+\delta J_{z}^{2},
\label{eq:2-2-16-b} \\
\delta J_{z}^{2} & =N\cot^{2}\theta\left[2n+N\cot^{2}\theta\right.  \notag \\
& \left.-\frac{(N+1)(n+N\cot^{2}\theta)}{(N-n+1)}\Gamma\left(n,N,\theta%
\right)\right],  \label{eq:2-2-16-c}
\end{align}
where
\end{subequations}
\begin{equation}
\left.\left\langle J_{z}^{2}\right\rangle \right|_{\theta=\frac{\pi}{2}%
}=\left(n-\frac{N}{2}\right)^{2}  \label{eq:2-2-17}
\end{equation}
is mean $z-$component spin square of the Dicke state $\left\vert
n-N/2\right\rangle $ and
\begin{equation}
\Gamma\left(n,N,\theta\right)=\frac{_{1}F_{1}(-n;N-n+2;-N\cot^{2}\theta)}{%
_{1}F_{1}(-n;N-n+1;-N\cot^{2}\theta)}.  \label{eq:2-2-18}
\end{equation}
For convenience we define equivalent squeezing parameter~\cite{Wang-1}
\begin{equation}
\zeta_{3}^{2}=\max\{1-\xi_{3}^{2},0\},  \label{eq:2-2-19}
\end{equation}
which characterizes the spin squeezing when $0<\zeta_{3}^{2}\leq1.$

%
\begin{figure}[ptb]
\begin{centering}
\includegraphics[bb=28 449 558 782,clip,width=3in]{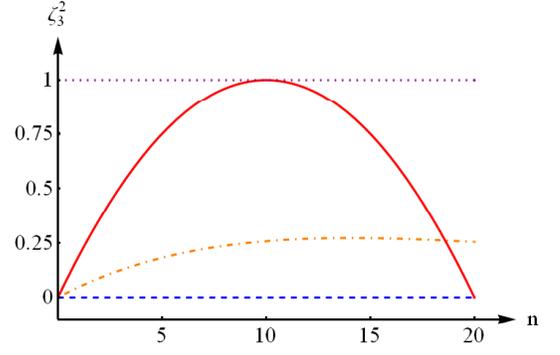}
\par\end{centering}
\caption{(Color online) The squeezing parameter $\protect\zeta_{3}^{2}$
versus the excitation number $n$ for $\protect\theta=0$ (blue dashed line), $%
\protect\theta=\protect\pi/4$ (orange dotdashed line) and $\protect\theta=%
\protect\pi/2$ (red solid line). The total number of the atoms $N$ are
chosen as $20$. The horizontal purple dotted line as the base line
represents the upper limit of the spin-squeezing. The dark state is always
squeezed except without excitation ($n=0$) or fully excited ($n=N$).}
\label{fig:fig2}
\end{figure}

%
\begin{figure}[ptb]
\begin{centering}
\includegraphics[bb=17 232 583 788,clip,width=3in]{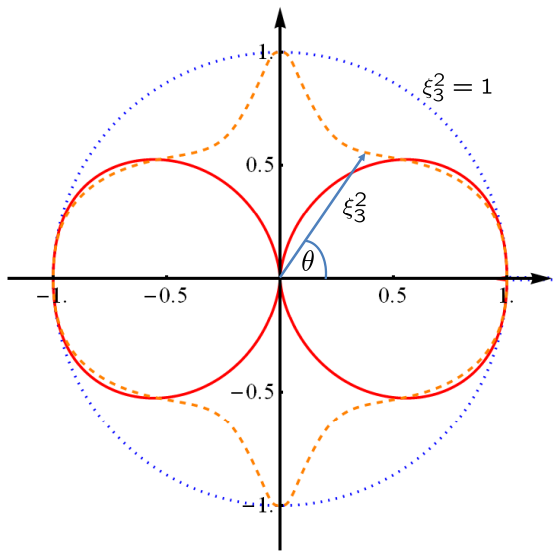}
\par\end{centering}
\caption{(Color online) The polar plot of the squeezing parameter $\protect%
\xi_{3}^{2}$ versus the parameter $\protect\theta$ for the excitation number
$n=0$ (blue dotted line), $n=10$ (orange dashed line) and $n=20$ (red solid
line). The total number of the atoms $N$ are chosen as $20$. The greatest
spin-squeezing is obtained for the symmetric Dicke state when the half of
the quasi-spins are excited ($n=N/2$) and $\protect\theta=\protect\pi/2$.}
\label{fig:fig3}
\end{figure}

%
\begin{figure}[ptb]
\begin{centering}
\includegraphics[bb=18 363 519 757,clip,width=3.2in]{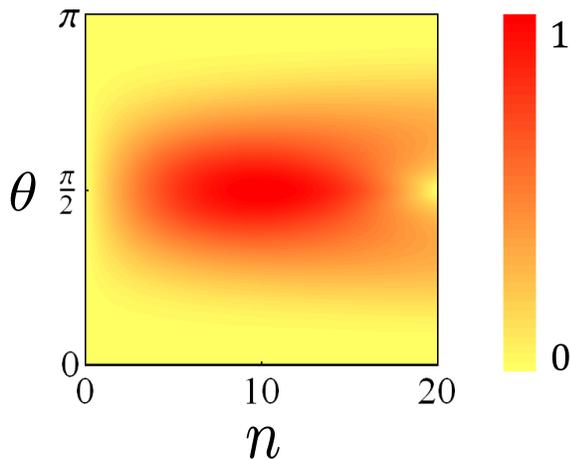}
\par\end{centering}
\caption{(Color online) The contour plot of the squeezing parameter $\protect%
\xi_{3}^{2}$ versus the excitation number $n$ and parameter $\protect\theta$%
. The total number of the atoms $N$ are chosen as $20$. The greatest
spin-squeezing ($\protect\zeta_{3}^{2}=1$) is obtained for the symmetric
Dicke state when the half of the quasi-spins are excited ($n=N/2$) and $%
\protect\theta=\protect\pi/2$.}
\label{fig:fig4}
\end{figure}


\subsection{Numerical pictures of the atomic squeezing in dark states}

The squeezing parameters of the dark state $\left\vert
d_{n}\left(\theta\right)\right\rangle $ without wave vector difference ($%
\mathbf{K}=0$) are illustrated in Fig.~\ref{fig:fig2}, ~\ref{fig:fig3} and ~%
\ref{fig:fig4}. In all three figures, the total number of the atoms $N$ are
chosen as $20$. Since the reduced density matrix in Eq. (\ref{eq:2-2-2}) is
symmetric when $\theta$ is replaced by $\theta+\pi$, the range of the $%
\theta $ is chosen as $\left[0,\pi\right)$. Fig.~\ref{fig:fig2} shows that
the squeezing parameter $\zeta_{3}^{2}$ varies with the excitation number $n$
for different $\theta$. The dark state is always squeezed except without
excitation ($n=0$) or fully excited ($n=N$). It is noticed that when $%
\theta\neq0,\pi/2$, the dark state is squeezed even when the excitation
number $n$ is equal to the total number of the atoms $N$. Because the dark
state actually mix the quasi-spins wave state and the photon Fock state,
this non-vanishing squeezing mainly results from the sub-Poisson
distribution of the photons, which will be explicitly demonstrated in the
last subsection of the Sec.~\ref{sec:two}. Fig.~\ref{fig:fig3} is the polar
plot of the squeezing parameter $\rho=\xi_{3}^{2}(\theta)$ versus $\theta$
for different excitation numbers $n$. For the state without excitation ($n=0$%
), the atomic ensemble always has no spin-squeezing because all of the
quasi-spins are in their ground states. The greatest spin-squeezing is
obtained for the symmetric Dicke state when the half of the quasi-spins are
excited ($n=N/2$) and $\theta=\pi/2$. Fig.~\ref{fig:fig4} is the contour
plot of the squeezing parameter $\zeta_{3}^{2}$ versus the excitation number
$n$ and parameter $\theta$. Obviously, the greatest spin-squeezing ($%
\zeta_{3}^{2}=1$ drawn by red color) can be obtained by adiabatically
manipulating the parameter $\theta$ to $\pi/2$ at the half excitation of the
quasi-spins ($n=N/2$).

%
\begin{figure}[ptb]
\begin{centering}
\includegraphics[bb=40 260 567 780,clip,width=3in]{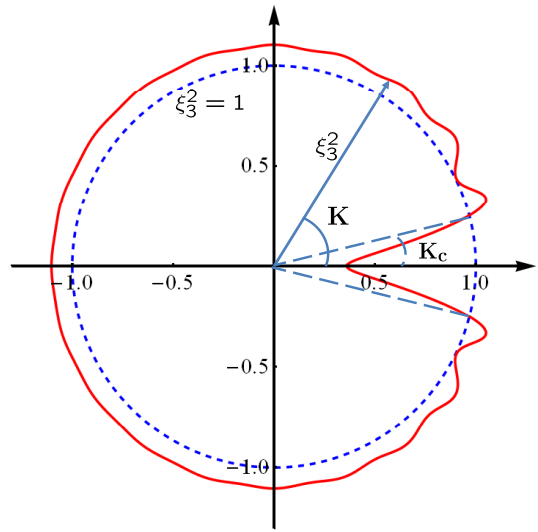}
\par\end{centering}
\caption{(Color online) The polar plot of the squeezing parameter $\protect%
\xi_{3}^{2}$ versus the vector difference $\mathbf{K}$. The total number of
the atoms $N$, the excitation number $n$ and the parameter $\protect\theta$
are chosen as $20,4$ and $\protect\pi/2$. Here, $\protect\xi_{3}^{2}=1$ is
the baseline to measure whether the dark state is squeezed or not.
Obviously, the dark state is squeezed only when the vector difference $%
\mathbf{K}\in[-\mathbf{K}_{\mathbf{c}},\mathbf{K}_{\mathbf{c}}]$. }
\label{fig:fig5}
\end{figure}


%
\begin{figure}[ptb]
\begin{centering}
\includegraphics[bb=34 335 517 724,clip,width=3.2in]{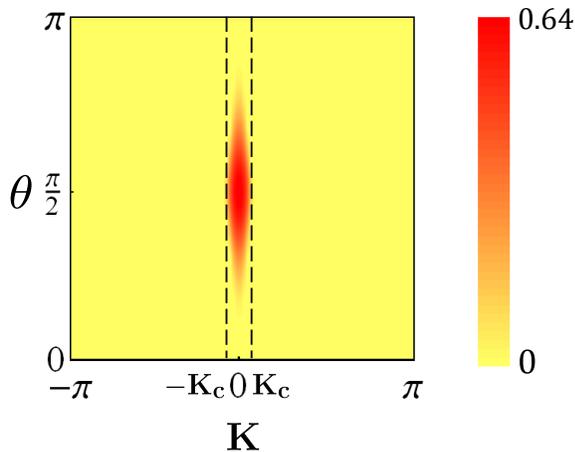}
\par\end{centering}
\caption{(Color online) The contour plot of the squeezing parameter $\protect%
\xi_{3}^{2}$ versus the vector difference $\mathbf{K}$ and the parameter $%
\protect\theta$. The total number of the atoms $N$ and the excitation number
$n$ are chosen as $20,4$. Obviously, the dark state is squeezed only when
the vector difference $\mathbf{K}\in[-\mathbf{K}_{\mathbf{c}},\mathbf{K}_{%
\mathbf{c}}]$. }
\label{fig:fig6}
\end{figure}


If the wave vectors between the quantum and the classical field are
different, the nonzero $\mathbf{K}$ results in the nonsymmetrical Dicke
bases in Eq. (\ref{eq:2-2-2}) and the spin squeezing decreases rapidly as
the $\mathbf{K}$ deviates from $0.$ Fig.~\ref{fig:fig5} is the polar plot of
the squeezing parameter $\rho=\xi_{3}^{2}(\theta)$ versus $\mathbf{K}$. The
parameters are chosen as $N=20,n=4,$ and $\theta=\pi/2$. Since $%
\xi_{3}^{2}=1 $ is the baseline to measure the atomic squeezing, the dark
state is squeezed only when the vector difference $\mathbf{K}\in[-\mathbf{K}%
_{\mathbf{c}},\mathbf{K}_{\mathbf{c}}]$, where $\mathbf{K}_{\mathbf{c}}$ is
shown in Fig.~\ref{fig:fig5}. Fig.~\ref{fig:fig6} is the contour plot of the
squeezing parameter $\zeta_{3}^{2}$ versus the vector difference $\mathbf{K}$
and parameter $\theta$. The total number of the atoms $N$ and the excitation
number $n$ are chosen as $20,4$. In Fig.~\ref{fig:fig6}, $[-\mathbf{K}_{%
\mathbf{c}},\mathbf{K}_{\mathbf{c}}]$ is still the boundary for the vector
difference $K$ when the dark states are squeezed.

\section{\label{sec:three}Internal entanglement in dark states and photon statistics}

\subsection{Concurrence in dark states without decoherence}

As the physical observable to quantify the pairwise entanglement of
spin-1/2, the concurrence is closely related to spin squeezing~\cite{Wang-2}%
. The concurrence is defined as~\cite{Wootters}
\begin{equation}
C=\max\left\{ 0,\lambda_{1}-\lambda_{2}-\lambda_{3}-\lambda_{4}\right\} ,
\label{eq:3-1-1}
\end{equation}
where the quantities $\lambda_{i}(i=1,2,3,4)$ are the square roots of the
eigen-values in descending order of the matrix product%
\begin{equation}
\varrho_{12}=\rho_{12}\left(\sigma_{1y}\otimes\sigma_{2y}\right)%
\rho_{12}^{*}\left(\sigma_{1y}\otimes\sigma_{2y}\right).  \label{eq:3-1-2}
\end{equation}
Here, $\rho_{12}$ is the two-spin reduced density matrix and $\rho_{12}^{*}$
is the complex conjugate of the $\rho_{12}$.

For the dark states, the two-quasi-spin reduced density matrix~\cite{Wang-3}
\begin{equation}
\rho_{12}=\left[%
\begin{array}{cccc}
v_{+} & 0 & 0 & u^{*} \\
0 & w & y & 0 \\
0 & y & w & 0 \\
u & 0 & 0 & v_{-}%
\end{array}%
\right]  \label{eq:3-1-3}
\end{equation}
can be explicitly obtained by tracing out the degrees of freedom of all the
other quasi-spins. It is clear to verify that for the dark states only the
elements $v_{\pm},u,w$ and $y$ of the $\rho_{12}$ survive as ($\phi=\mathbf{%
K\cdot l}$)
\begin{subequations}
\begin{align}
v_{\pm} & =\left.v_{\pm}\right|_{\theta=\frac{\pi}{2}}+\delta v_{\pm},
\label{eq:3-1-4-a} \\
w & =\left.w\right|_{\theta=\frac{\pi}{2}}+\delta w,  \label{eq:3-1-4-b} \\
u & =-wi\sin\mathbf{\phi},  \label{eq:3-1-4-c} \\
y & =w\cos\phi,  \label{eq:3-1-4-d}
\end{align}
where~\cite{Wang-3}
\end{subequations}
\begin{subequations}
\begin{align}
\left.v_{+}\right|_{\theta=\frac{\pi}{2}} & =\frac{1}{4}\left(1+2\left%
\langle \sigma_{z}^{1}\right\rangle +\left\langle
\sigma_{z}^{1}\sigma_{z}^{2}\right\rangle \right)  \notag \\
& =\frac{n(n-1)}{N(N-1)},  \label{eq:3-1-5-a} \\
\left.v_{-}\right|_{\theta=\frac{\pi}{2}} & =\frac{1}{4}\left(1-2\left%
\langle \sigma_{z}^{1}\right\rangle +\left\langle
\sigma_{z}^{1}\sigma_{z}^{2}\right\rangle \right)  \notag \\
& =\frac{(n-N)(n-N+1)}{N(N-1)},  \label{eq:3-1-5-b} \\
\left.w\right|_{\theta=\frac{\pi}{2}} & =\frac{1}{4}\left(1-\left\langle
\sigma_{z}^{1}\sigma_{z}^{2}\right\rangle \right)  \notag \\
& =\frac{nN-n^{2}}{N(N-1)}  \label{eq:3-1-5-c}
\end{align}
are the corresponding elements of the reduced density matrix over the Dicke
state $\left\vert n-N/2\right\rangle $ and
\end{subequations}
\begin{subequations}
\begin{align}
\delta v_{+} & =B\left(N,\theta\right)\left[n+(N\csc^{2}\theta+n-1)%
\Lambda(n,N,\theta)\right],  \label{eq:3-1-6-a} \\
\delta v_{-} & =B\left(N,\theta\right)\left[n+(N\cot^{2}\theta-N+2n+1)%
\Lambda(n,N,\theta)\right],  \label{eq:3-1-6-b} \\
\delta w & =B\left(N,\theta\right)\left[-n-\left(n+N\cot^{2}\theta\right)%
\Lambda(n,N,\theta)\right]  \label{eq:3-1-6-c}
\end{align}
with $B\left(N,\theta\right)=\cot^{2}\theta/\left(N-1\right)$ and
\end{subequations}
\begin{equation}
\Lambda(n,N,\theta)=1-\frac{(N+1)}{(N-n+1)}\Gamma\left(n,N,\theta\right),
\label{eq:3-1-7}
\end{equation}
where $\Gamma\left(n,N,\theta\right)$ is defined in Eq. (\ref{eq:2-2-18}).

\begin{figure}[ptb]
\begin{centering}
\includegraphics[bb=23 394 541 773,clip,width=3.2in]{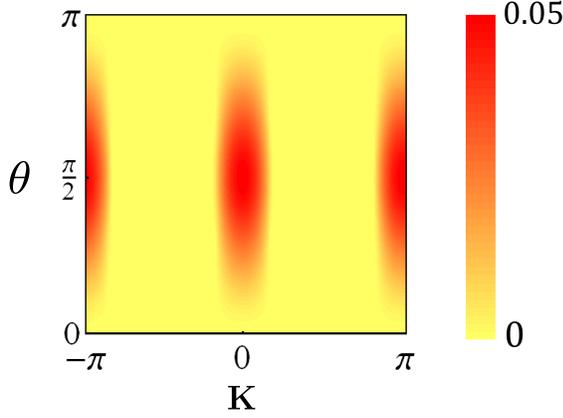}
\par\end{centering}
\caption{(Color online) The contour plot of the concurrence $C$ versus the
vector difference $\mathbf{K}$ and the parameter $\protect\theta$. The total
number of the atoms $N$ and the excitation number $n$ are chosen as $20$ and
$4$. Here we find three regions for generating concurrence rather than one
region for generating atomic squeezing.}
\label{fig:fig7}
\end{figure}

\begin{figure}[ptb]
\begin{centering}
\includegraphics[bb=42 363 565 707,clip,width=3in]{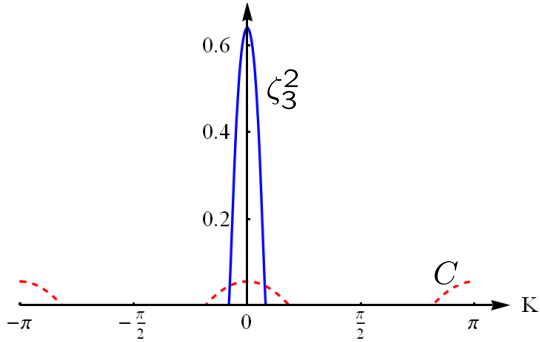}
\par\end{centering}
\caption{(Color online) The sections of the atomic squeezing $\protect\zeta%
_{3}^{2}$ (blue solid blue) and concurrence $C$ (red dashed line) versus the
vector difference $\mathbf{K}$. The total number of the atoms $N$, the
excitation number $n$ and the parameter $\protect\theta$ are chosen as $20,4$
and $\protect\pi/2$. Obviously, in the vicinity of the $\mathbf{K}=\protect%
\pi$, the concurrence appears but atomic squeezing does not.}
\label{fig:fig8}
\end{figure}

Thus the concurrence of the dark state is given by
\begin{equation}
C=\max\left\{ 0,2\left(w\left\vert \cos\phi\right\vert -\sqrt{v_{+}v_{-}}%
\right)\right\} .  \label{eq:3-1-8}
\end{equation}
Fig.~\ref{fig:fig7} is the contour plot of the concurrence $C$ versus the
vector difference $\mathbf{K}$ and the parameter $\theta$. In comparison
with the one region $[-\mathbf{K}_{\mathbf{c}},\mathbf{K}_{\mathbf{c}}]$ for
generating atomic squeezing, there are three regions for generating pairwise
entanglement. This is also shown in Fig.~\ref{fig:fig8}, which is the
sections of the atomic squeezing and the concurrence. Obviously, in the
vicinity of the $K=\pi$, the concurrence appears but atomic squeezing does
not. In Fig.~\ref{fig:fig7} and \ref{fig:fig8}, the total number of the
atoms $N$ and the excitation number $n$ are chosen as $20,4$. In the next
section, the evolvement of the concurrence as well as the spin-squeezing
under various decoherence channels will be considered.

%

%

%
\begin{figure}[ptb]
\begin{centering}
\includegraphics[bb=21 476 549 777,clip,width=3in]{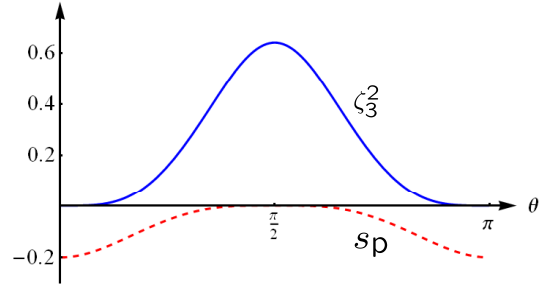}
\par\end{centering}
\caption{(Color online) The sections of the atomic squeezing $\protect\zeta%
_{3}^{2}$ (blue solid blue) and sub-Poisson distribution $s_{\mathbf{p}}$
(red dashed line) versus the parameter $\protect\theta$. The total number of
the atoms $N$, the excitation number $n$ and the vector difference $K$ are
chosen as $20,4$ and $0$. When the $\protect\theta$ increases from $0$ to $%
\protect\pi/2$, the atomic squeezing increases and the sub-Poisson
distribution decreases simultaneously and vise vesa when the $\protect\theta$
increases from $\protect\pi/2$ to $\protect\pi$.}
\label{fig:fig9}
\end{figure}


\subsection{Sub-Poisson distribution of the photons}

We have shown that the atomic squeezing for information storage can be
generated through the adiabatically manipulating the angle $\theta$. In
another aspect, as the photon dressed collective atom state the dark states
mix the atomic ensemble state and the photon states. In our setup, the
conservative quantity is the summation of the $z-$component collective
operator $J_{z}$ and the number operator $n=a^{\dagger}a$ of the photons.
The multipartite quantum correlation in dark states actually is the $z-$%
component spin squeezing for the dark states, which would result from the
quantum correlation of the photons. Usually, such photonic quantum
correlation can be measured by the sub-Poisson distribution of the photons
defined as%
\begin{align}
s_{\mathrm{p}} & =\frac{1}{N}\left(\left\langle \Delta n^{2}\right\rangle
-\left\langle n\right\rangle \right)  \notag \\
& =\frac{1}{N}\left(\left\langle a^{\dagger}a^{\dagger}aa\right\rangle
-\left\langle a^{\dagger}a\right\rangle ^{2}\right).  \label{eq:3-2-1}
\end{align}
If the photonic number fluctuation
\begin{equation}
\left\langle \Delta n^{2}\right\rangle =\left\langle n^{2}\right\rangle
-\left\langle n\right\rangle ^{2}  \label{eq:3-2-2}
\end{equation}
is smaller than the average of the particle number $\left\langle
n\right\rangle $ of the photons ($s_{\mathrm{p}}<0$), the dark states
actually has the sub-Poisson distribution.

The straightforward calculation gives
\begin{subequations}
\begin{align}
\left\langle a^{\dagger}a^{\dagger}aa\right\rangle & =\frac{%
\left(n-1\right)nN^{2}\cot^{4}\theta}{(N-n+2)(N-n+1)}\times  \notag \\
& \frac{_{1}F_{1}(2-n;N-n+3;-N\cot^{2}\theta)}{_{1}F_{1}(-n;N-n+1;-N\cot^{2}%
\theta)},  \label{eq:3-2-3-a} \\
\left\langle a^{\dagger}a\right\rangle & =\frac{nN\cot^{2}\theta}{(N-n+1)}%
\times  \notag \\
& \frac{_{1}F_{1}(1-n;N-n+2;-N\cot^{2}\theta)}{_{1}F_{1}(-n;N-n+1;-N\cot^{2}%
\theta)}.  \label{eq:3-2-3-b}
\end{align}
Substituting the relevant expectation values into Eqs. (\ref{eq:3-2-1})
leads to the explicit expression of the sub-Poisson distribution of the
photons $s_{\mathrm{p}}$, which always follow the atomic squeezing $%
\zeta_{3}^{2}$. Fig.~\ref{fig:fig9} illustrates the numerical relationship
between sub-Poisson distribution and the atomic squeezing. When the $\theta$
increases from $0$ to $\pi/2$, the atomic squeezing increases and the
sub-Poisson distribution decreases simultaneously and vise vesa when the $%
\theta$ increases from $\pi/2$ to $\pi$. From the above discussion, the
greatest atomic squeezing and the concurrence can be generated when the
parameter $\theta$ is manipulated to $\pi/2$, which means all the
sub-Poisson distribution of the photons converts to the atomic squeezing.

In this section, we just consider the adiabatic manipulation without
any decoherence process. In the next section, we will take the
typical three atomic decoherence channels into consideration to
estimate the sudden death of the atomic squeezing and concurrence.
Since the coherence time of photons is much larger than atomic one,
the photonic decoherence such as the dephasing due to the imperfect
single-photon indistinguishability~\cite{photonic} is not considered
below.

\section{~\label{sec:four}Atomic Squeezing And Pairwise Entanglement Under Decoherence
Channels}

\subsection{Decoherence channels}

Through the adiabatically manipulating the angle $\theta$, the
spin-squeezing and the concurrence are generated for information storage.
However, due to the existence of the environment, they are irreversible
depressed and vanish eventually. Such decoherence processes are usually
described by three typical decoherence channels: the amplitude damping
channel (ADC), the phase damping channel (PDC) and the depolarizing channel
(DPC)~\cite{Wang-2,decoherence}. They are defined by some maps of the
elements of the density matrix $\left|i\right\rangle \left\langle j\right|$.
Since the ADC is an energy-losing process and thus it forces the initial
state to dissipate into the ground state, the map is defined as
\end{subequations}
\begin{subequations}
\begin{align}
\left|i\right\rangle \left\langle i\right| &
\rightarrow(1-p)\left|i\right\rangle \left\langle
i\right|+p\left|0\right\rangle \left\langle 0\right|,  \label{eq:4-1-1-a} \\
\left|i\right\rangle \left\langle j\right| & \rightarrow(1-p)^{\frac{1}{2}%
}\left|i\right\rangle \left\langle j\right|,(i\neq j).  \label{eq:4-1-1-b}
\end{align}
The PDC is phase-losing process and the corresponding map is described by
\end{subequations}
\begin{equation}
\left|i\right\rangle \left\langle
j\right|\rightarrow\left(1-p+p\delta_{i,j}\right)\left|i\right\rangle
\left\langle j\right|.  \label{eq:4-1-2}
\end{equation}
The DPC is polarization-losing process and the corresponding map is
represented by
\begin{equation}
\left|i\right\rangle \left\langle
j\right|\rightarrow\left(1-p\right)\left|i\right\rangle \left\langle
j\right|+p\delta_{i,j}\frac{\mathbf{I}}{2}.  \label{eq:4-1-3}
\end{equation}
The behaviors of the spin-squeezing and the concurrence under these
decoherence channels will be present in the following sub sections. Here, $p$
is the decoherence strength whose range is $\left[0,1\right].$ When $p=0$
there is no decoherence and when $p=1$ the decoherence processes are
completed.

If we fix the initial values of the spin-squeezing and the concurrence of
the dark state, the dark state degrades to a product state with single
quasi-spin state as a component when the decoherence strength $p$ increases
from $0$. Thus the spin-squeezing and the concurrence would have sudden
death. Before the sudden death happens, the information storage can be in
progress safely. In other word, the longer time the spin-squeezing and the
concurrence survive, the better the information storage can be achieved.

From the above expression of the spin-squeezing and the concurrence, we
notice that if we know the expectations $\left\langle J_{z}\right\rangle ,$$%
\left\langle J_{z}^{2}\right\rangle $and $\left\langle
\sigma_{z}^{1}\right\rangle $, and the correlations $\left\langle
\sigma_{z}^{1}\sigma_{z}^{2}\right\rangle ,$ all the spin-squeezing and the
concurrence can be determined. We will give explicit analytical expressions
for them under three decoherence channels.

\subsection{Amplitude-damping channel}

Based on the map in Eq. (43) for ADC, one can find the following relations
for single quasi-spin operators~\cite{Wang-1}
\begin{subequations}
\begin{eqnarray}
\left\langle \sigma_{z}^{1}\left(p\right)\right\rangle & = & s\left\langle
\sigma_{z}^{1}\right\rangle _{0}-p,  \label{eq:4-2-1-a} \\
\left\langle \sigma_{z}^{1}\sigma_{z}^{2}\left(p\right)\right\rangle & = &
s^{2}\left\langle \sigma_{z}^{1}\sigma_{z}^{2}\right\rangle
_{0}-2sp\left\langle \sigma_{z}^{1}\right\rangle _{0}+p^{2},
\label{eq:4-2-1-b} \\
\left\langle
\sigma_{\alpha}^{1}\sigma_{\beta}^{2}\left(p\right)\right\rangle & = &
s\left\langle \sigma_{\alpha}^{1}\sigma_{\beta}^{2}\right\rangle
_{0}(\alpha,\beta=x,y).  \label{eq:4-2-1-c}
\end{eqnarray}
Hereafter, we define $s=1-p$. Then the corresponding relations for the
collective operators are
\end{subequations}
\begin{subequations}
\begin{eqnarray}
\left\langle J_{z}^{2}\left(p\right)\right\rangle & = & s^{2}\left\langle
J_{z}^{2}\right\rangle _{0}+(N-1)sp\left\langle J_{z}\right\rangle _{0}
\notag \\
& & +\frac{N^{2}}{4}p\left(p+\frac{2}{N}s\right),  \label{eq:4-2-2-a} \\
\left\langle J_{z}\left(p\right)\right\rangle & = & s\left\langle
J_{z}\right\rangle _{0}-\frac{N}{2}p,  \label{eq:4-2-2-b} \\
\left\langle \mathbf{J}^{2}\left(p\right)\right\rangle & = & s\left\langle
\mathbf{J}^{2}\right\rangle _{0}-sp\left\langle J_{z}^{2}\right\rangle
_{0}+(N-1)sp\left\langle J_{z}\right\rangle _{0}  \notag \\
& & +\frac{N}{4}p(Np+2s).  \label{eq:4-2-2-c}
\end{eqnarray}
Substituting the relevant expectation values and the correlation function
into Eqs. (\ref{eq:2-2-13},\ref{eq:2-2-14}) leads to the explicit expression
of the spin-squeezing parameters
\end{subequations}
\begin{align}
\xi_{3}^{2}\left(p\right)^{\mathrm{A}} & =\frac{\varsigma^{2}\left(p\right)}{%
\frac{4}{N^{2}}\left\langle \mathbf{J}^{2}\left(p\right)\right\rangle -\frac{%
2}{N}}  \label{eq:4-2-3}
\end{align}
with
\begin{equation}
\varsigma^{2}\left(p\right)=s^{2}\varsigma_{0}^{2}+2\frac{4}{N}%
(N-1)sp\left\langle J_{z}\right\rangle _{0}+p(1+s),  \label{eq:4-2-4}
\end{equation}
where $\varsigma_{0}^{2}$ is the initial spin-squeezing parameter defined in
Eq. (\ref{eq:2-2-13}).

In the same case, in order to investigate the evolution of the concurrence,
the two-qubit density matrix is shown as
\begin{equation}
\rho_{12}^{\mathrm{A}}=\left[%
\begin{array}{cccc}
v_{+}^{\mathrm{A}} & 0 & 0 & u^{\mathrm{A},*} \\
0 & w^{\mathrm{A}} & y^{\mathrm{A}} & 0 \\
0 & y^{\mathrm{A}} & w^{\mathrm{A}} & 0 \\
u^{\mathrm{A}} & 0 & 0 & v_{-}^{\mathrm{A}}%
\end{array}%
\right]  \label{eq:4-2-5}
\end{equation}
with $v_{+}^{\mathrm{A}}=s^{2}v_{+},$ $y^{\mathrm{A}}=sy,$ $u^{\mathrm{A}%
}=su $ and
\begin{subequations}
\begin{align}
v_{-}^{\mathrm{A}} & =s^{2}v_{+}-s\left\langle \sigma_{z}\right\rangle
_{0}+p,  \label{eq:4-2-6-a} \\
w^{\mathrm{A}} & =s^{2}w+\frac{1}{2}sp\left\langle
\sigma_{z}^{1}\right\rangle _{0}+\frac{1}{2}sp,  \label{eq:4-2-6-b}
\end{align}
where $v_{\pm},y,u$ and $w$ are initial elements defined in Eq. (35). Thus
we obtain the time evolution of the concurrence as
\end{subequations}
\begin{subequations}
\begin{align}
C^{\mathrm{A}} & =\max\left\{ 0,C_{1}^{\mathrm{A}},C_{2}^{\mathrm{A}%
}\right\} ,  \label{eq:4-2-7-a} \\
C_{1}^{\mathrm{A}} & =2\left(sw\left\vert \cos\phi\right\vert -\sqrt{v_{+}^{%
\mathrm{A}}v_{-}^{\mathrm{A}}}\right),  \label{eq:4-2-7-b} \\
C_{2}^{\mathrm{A}} & =2\left(sw\left\vert \sin\phi\right\vert -w^{\mathrm{A}%
}\right).  \label{eq:4-2-7-c}
\end{align}
Hereafter, we only consider the case at the vector difference $\mathrm{K=0}$
because the quantum correlations decrease rapidly when the vector difference
deviates from $0$ shown in Fig.~\ref{fig:fig5}.

The subindex $\mathrm{A}$ in the above Eqs. (\ref{eq:4-2-3},\ref{eq:4-2-5}%
,51,52) represents that the evolutions of the spin-squeezing and the
concurrence are taken under ADC. The sudden death of the spin-squeezing and
the concurrence implied in Eq. (\ref{eq:4-2-3},52) are shown in Fig.~\ref%
{fig:fig10}. Usually, the atomic squeezing $\zeta_{3}^{2}$ disappear later
than the concurrence $C$ as shown in Fig.~\ref{fig:fig10}(a). However for
ADC, the atomic squeezing $\zeta_{3}^{2}$ can disappear earlier than the
concurrence $C$ (Fig.~\ref{fig:fig10}(b)). With appropriate parameters, the
atomic squeezing $\zeta_{3}^{2}$ may reach its maximum and then disappear
near the maximum decoherence strength $p=1$ (Fig.~\ref{fig:fig10}(c)). The
parameters are chosen as $N=20,n=16,K=0$ and (a)$\theta=\pi/3$,(b)$%
\theta=\pi/2$,(c)$\theta=0.673\pi$.

%
\begin{figure}[ptb]
\begin{centering}
\includegraphics[bb=26 401 542 773,clip,width=3.5in]{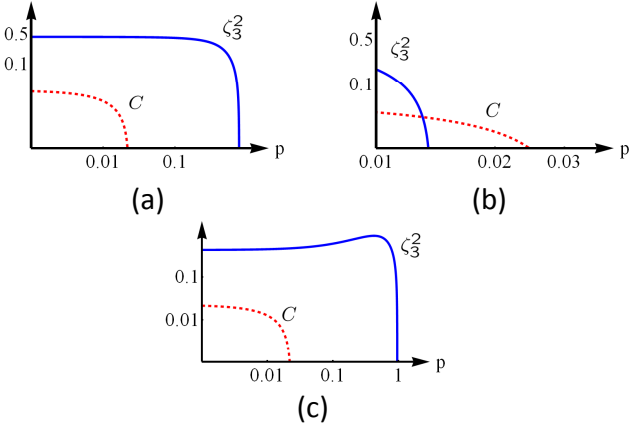}
\par\end{centering}
\caption{(Color online) The sudden death of the atomic squeezing $\protect%
\zeta_{3}^{2}$ (blue solid line) and the concurrence $C$ (red dashed line)
under ADC. (a)Usually, the atomic squeezing $\protect\zeta_{3}^{2}$
disappear later than the concurrence $C$. (b)The atomic squeezing $\protect%
\zeta_{3}^{2}$ can disappear earlier than the concurrence $C$. (c)The atomic
squeezing $\protect\zeta_{3}^{2}$ may reach its maximum and then disappear
near the maximum decoherence strength $p=1$. The parameters are chosen as $%
N=20,n=16,K=0$ and (a)$\protect\theta=\protect\pi/3$,(b)$\protect\theta=%
\protect\pi/2$,(c)$\protect\theta=0.673\protect\pi$.}
\label{fig:fig10}
\end{figure}


\subsection{Phase-damping channel}

For PDC described by the map in Eq. (\ref{eq:4-1-2}), one can find the
following relations for single quasi-spin operators~\cite{Wang-1}
\end{subequations}
\begin{subequations}
\begin{eqnarray}
\left\langle \sigma_{z}\left(p\right)\right\rangle & = & \left\langle
\sigma_{z}\right\rangle _{0},  \label{eq:4-3-1-a} \\
\left\langle \sigma_{z}^{1}\sigma_{z}^{2}\left(p\right)\right\rangle & = &
\left\langle \sigma_{z}^{1}\sigma_{z}^{2}\right\rangle _{0},
\label{eq:4-3-1-b} \\
\left\langle
\sigma_{\alpha}^{1}\sigma_{\beta}^{2}\left(p\right)\right\rangle & = &
s^{2}\left\langle \sigma_{\alpha}^{1}\sigma_{\beta}^{2}\right\rangle
_{0}(\alpha,\beta=x,y).  \label{eq:4-3-1-c}
\end{eqnarray}
Then the corresponding relations for the collective operators are $%
\left\langle J_{z}^{2}\left(p\right)\right\rangle =\left\langle
J_{z}^{2}\right\rangle _{0},\left\langle J_{z}\left(p\right)\right\rangle
=\left\langle J_{z}\right\rangle _{0}$ and
\end{subequations}
\begin{eqnarray}
\left\langle \mathbf{J}^{2}\left(p\right)\right\rangle & = &
s^{2}\left\langle \mathbf{J}^{2}\right\rangle _{0}+(1-s^{2})\frac{N}{2}.
\label{eq:4-3-2}
\end{eqnarray}
Substituting the relevant expectation values and the correlation function
into Eqs. (\ref{eq:2-2-13},\ref{eq:2-2-14}) leads to the explicit expression
of the spin-squeezing parameters%
\begin{align}
\xi_{3}^{2}\left(p\right)^{\mathrm{P}} & =\frac{\varsigma_{0}^{2}}{\frac{4}{%
N^{2}}\left\langle \mathbf{J}^{2}\left(p\right)\right\rangle -\frac{2}{N}}.
\label{eq:4-3-3}
\end{align}
In the same sense to investigate the evolution of the concurrence, the
two-qubit density matrix is present as%
\begin{equation}
\rho_{12}^{\mathrm{P}}=\left[%
\begin{array}{cccc}
v_{+}^{\mathrm{P}} & 0 & 0 & u^{\mathrm{P},*} \\
0 & w^{\mathrm{P}} & y^{\mathrm{P}} & 0 \\
0 & y^{\mathrm{P}} & w^{\mathrm{P}} & 0 \\
u^{\mathrm{P}} & 0 & 0 & v_{-}^{\mathrm{P}}%
\end{array}%
\right]  \label{eq:4-3-4}
\end{equation}
with relations $v_{\pm}^{\mathrm{P}}=v_{\pm},$ $w^{\mathrm{P}}=w,$ $y^{%
\mathrm{P}}=s^{2}y,$ and $u^{\mathrm{P}}=s^{2}u$. Thus we obtain the
evolution of the concurrence as%
\begin{equation}
C^{\mathrm{\mathrm{P}}}=\max\left\{ 0,2\left(s^{2}w\left\vert
\cos\phi\right\vert -\sqrt{v_{+}v_{-}}\right)\right\} .  \label{eq:4-3-5}
\end{equation}

The subindex $\mathrm{P}$ in the above Eqs. (\ref{eq:4-3-3},\ref{eq:4-3-4},%
\ref{eq:4-3-5}) represents that the time evolutions of the spin-squeezing
and the concurrence are taken under PDC. The sudden death of the
spin-squeezing and the concurrence implied in Eq. (\ref{eq:4-3-3},\ref%
{eq:4-3-5}) are shown in Fig. 11. In contrast of the ADC case, the atomic
squeezing $\zeta_{3}^{2}$ always disappear later than the concurrence $C$.

%
\begin{figure}[ptb]
\begin{centering}
\includegraphics[bb=23 517 568 754,clip,width=3.5in]{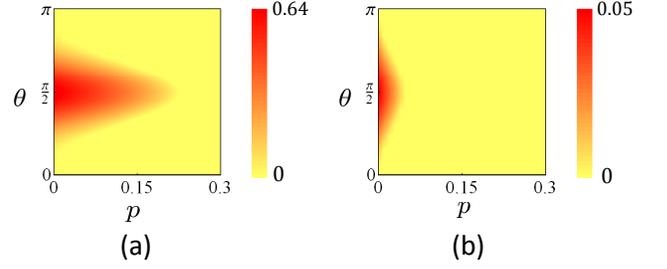}
\par\end{centering}
\caption{(Color online) The sudden death of (a) the atomic squeezing $%
\protect\zeta_{3}^{2}$ and (b) the concurrence $C$ under PDC. The parameters
are chosen as $N=20,n=16$ and $K=0$. The atomic squeezing $\protect\zeta%
_{3}^{2}$ always disappear later than the concurrence $C$.}
\label{fig:fig11}
\end{figure}

%
\begin{figure}[ptb]
\begin{centering}
\includegraphics[bb=19 501 570 745,clip,width=3.5in]{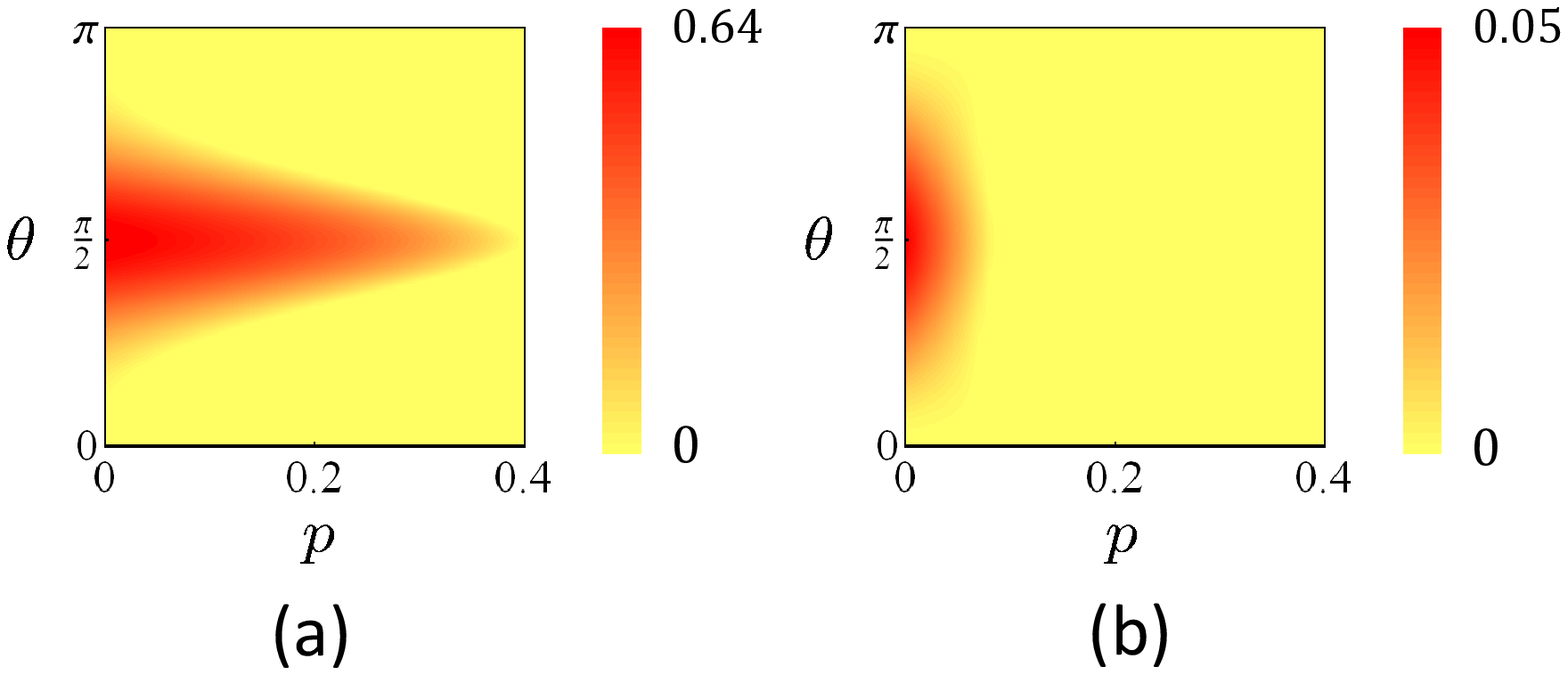}
\par\end{centering}
\caption{(Color online) The sudden death of (a) the atomic squeezing $%
\protect\zeta_{3}^{2}$ and (b) the concurrence $C$ under DPC. The parameters
are chosen as $N=20,n=4$ and $K=0$. The atomic squeezing $\protect\zeta%
_{3}^{2}$ always disappear later than the concurrence $C$.}
\label{fig:fig12}
\end{figure}


\subsection{Depolarizing channel}

Based on the map in Eq. (\ref{eq:4-1-3}) for DPC, one can find the following
relations for single quasi-spin operators~\cite{Wang-1}

\begin{subequations}
\begin{eqnarray}
\left\langle \sigma_{z}\left(p\right)\right\rangle & = & s\left\langle
\sigma_{z}\right\rangle _{0},  \label{eq:4-4-1-a} \\
\left\langle
\sigma_{\alpha}^{1}\sigma_{\beta}^{2}\left(p\right)\right\rangle & = &
s^{2}\left\langle \sigma_{\alpha}^{1}\sigma_{\beta}^{2}\right\rangle
_{0}(\alpha,\beta=x,y,z).  \label{eq:4-4-1-b}
\end{eqnarray}
Then the corresponding relations for the collective operators are $%
\left\langle J_{z}\left(p\right)\right\rangle =s\left\langle
J_{z}\right\rangle _{0}$ and
\end{subequations}
\begin{subequations}
\begin{eqnarray}
\left\langle J_{z}^{2}\left(p\right)\right\rangle & = & s^{2}\left\langle
J_{z}^{2}\right\rangle _{0}+\left(1-s^{2}\right)\frac{N}{4},
\label{eq:4-4-2-a} \\
\left\langle \mathbf{J}^{2}\left(p\right)\right\rangle & = &
s^{2}\left\langle \mathbf{J}^{2}\right\rangle _{0}+(1-s^{2})\frac{3N}{4}.
\label{eq:4-4-2-b}
\end{eqnarray}
Substituting the relevant expectation values and the correlation function
into Eqs. (\ref{eq:2-2-13},\ref{eq:2-2-14}) leads to the explicit expression
of the spin-squeezing parameters
\end{subequations}
\begin{align}
\xi_{3}^{2}\left(p\right)^{\mathrm{D}} & =\frac{s^{2}%
\varsigma_{0}^{2}+1-s^{2}}{\frac{4}{N^{2}}\left\langle \mathbf{J}%
^{2}\left(p\right)\right\rangle -\frac{2}{N}}.  \label{eq:4-4-3}
\end{align}
In the same sense to investigate the evolution of the concurrence, the
two-qubit density matrix is present as%
\begin{equation}
\rho_{12}^{\mathrm{D}}=\left[%
\begin{array}{cccc}
v_{+}^{\mathrm{D}} & 0 & 0 & u^{\mathrm{D},*} \\
0 & w^{\mathrm{D}} & y^{\mathrm{D}} & 0 \\
0 & y^{\mathrm{D}} & w^{\mathrm{D}} & 0 \\
u^{\mathrm{D}} & 0 & 0 & v_{-}^{\mathrm{D}}%
\end{array}%
\right]  \label{eq:4-4-5}
\end{equation}
with relations $y^{\mathrm{D}}=s^{2}y,$ $u^{\mathrm{D}}=s^{2}u$ and
\begin{subequations}
\begin{align}
v_{\pm}^{\mathrm{D}} & =\frac{s^{2}+s}{2}v_{\pm}+\frac{s^{2}-s}{2}v_{\mp}+%
\frac{1-s^{2}}{4},  \label{eq:4-4-6-a} \\
w^{\mathrm{D}} & =s^{2}w_{0}+\frac{1-s^{2}}{4}.  \label{eq:4-4-6-b}
\end{align}
Thus we obtain the evolution of the concurrence as
\end{subequations}
\begin{equation}
C^{\mathrm{\mathrm{D}}}=\max\left\{ 0,2\left(s^{2}w\left\vert
\cos\phi\right\vert -\sqrt{v_{+}^{\mathrm{D}}v_{-}^{\mathrm{D}}}%
\right)\right\} .  \label{eq:4-4-7}
\end{equation}

The subindex $\mathrm{D}$ in the above Eqs. (\ref{eq:4-4-3},\ref{eq:4-4-5}%
,63,\ref{eq:4-4-7}) represents that the evolutions of the spin-squeezing and
the concurrence are taken under DPC. The sudden death of the spin-squeezing
and the concurrence implied in Eq. (\ref{eq:4-4-3},\ref{eq:4-4-7}) are shown
in Fig.~\ref{fig:fig12}. In contrast of the ADC case, the atomic squeezing $%
\zeta_{3}^{2}$ always disappear later than the concurrence $C$.

\section{\label{sec:five}Optimal Time for Generating Atomic Squeezing}

Now, two competitive processes dominate generating the spin-squeezing. One
is adiabatically manipulating the parameter $\theta $ from $0$ to $\pi /2$,
which increases the spin-squeezing and the concurrence for storing the
information of phonons into the atomic ensemble. However, the other process
resulting from the various decoherence channels decreases the spin-squeezing
and the concurrence, which means that the system loses information
continuously. Therefore, the competition between this two processes leads to
the existence of an optimal time for storing information, after which the
information stored in the atomic ensemble is always losing. We can define
two time scales: if only considering the adiabatic manipulation, one time
scale is $t_{1}$ representing the time when the spin squeezing increases to
the half maximum value; if only considering the decoherence processes, the
other time scale is $t_{2}$ representing the time when the spin squeezing
decreases to the half maximum value. Only when $t_{1}>t_{2}$ the optimal
time exists.

We would like to demonstrate such competition in a typical quantum memory
based on the cold $^{87}\mathrm{Rb}$ atomic ensemble, which is released from
a magneto-optical trap at a temperature of about $100\mu K$~\cite{Pan}. In
such quantum memory, the ground state $\left\vert g\right\rangle $, the
metastable state $\left\vert m\right\rangle $ and the excited state $%
\left\vert e\right\rangle $ are chosen as $\left\vert
5S_{1/2},F=1,m_{F}=1\right\rangle $, $\left\vert
5S_{1/2},F=2,m_{F}=-1\right\rangle $ and $\left\vert
5S_{1/2},F=2,m_{F}=0\right\rangle $, respectively. An off-resonant $\sigma
^{-}$-polarized write pulse contributes to the transition from $\left\vert
g\right\rangle $ to $\left\vert e\right\rangle $ and the Stokes photon with $%
\sigma ^{-}$ polarization associates with the transition from
$\left\vert e\right\rangle $ to $\left\vert m\right\rangle $. The
dark state for storing the photonic information is described in Eq.
(\ref{eq:2-1-5}).

%
\begin{figure}[tbp]
\begin{centering}
\includegraphics[bb=41 482 519 772,clip,width=2.8in]{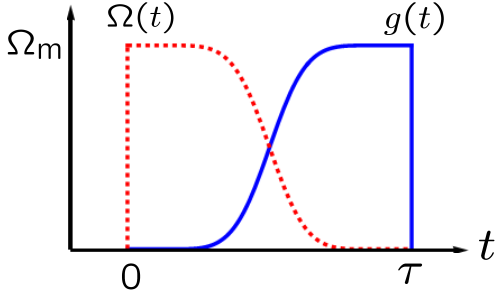}
\par\end{centering}
\caption{(Color online) Optical pulse sequences for both Rabi
frequencies. The red dashed line and black solid line represent the
Rabi frequencies $\Omega(t)$ and $g(t)$, respectively. The adiabatic
following condition is $\Omega _{m}\tau \gg 1.$} \label{fig:fig13}
\end{figure}

To realize the adiabatic creation of the non-classical correlation, we adopt
the hyperbolic tangent type pulse sequence for both Rabi frequencies~\cite%
{Rabi} as
\begin{subequations}
\label{5-0-1}
\begin{eqnarray}
g\left( t\right) &=&\Omega _{m}\left[ 1-\tanh (\frac{a}{t}+\frac{a}{t-\tau })%
\right] ,  \label{5-0-1-1} \\
\Omega \left( t\right) &=&\Omega _{m}\left[ 1+\tanh (\frac{a}{t}+\frac{a}{%
t-\tau })\right]  \label{5-0-1-2}
\end{eqnarray}%
instead of the usual gaussian type ones~\cite{Gaussian} or solitonary type ones~%
\cite{solitonary}. Here, $\Omega _{m}$ is the maximum Rabi
frequency, $\tau $ is the pulse length and $a$ corresponds to the
half width of the pulse. The pulse sequences of both Rabi
frequencies are depicted in Fig.~\ref{fig:fig13}. Only at the time
interval $\left[ 0,\tau \right] $ are both sequences applied to
adiabatically manipulate the dark state in Eq. (\ref{eq:2-1-5}). Two
Rabi frequencies respectively reach their maximum value $\Omega
_{m}$ at the initial time $t=0$ and the terminal time $t=\tau $,
which guarantees that the angle $\theta $ varies from $0$ to $\pi
/2$ and thus the information of the photons is stored into the
atomic ensemble at the time interval $\left[ 0,\tau \right] $.

Actually, due to the adiabatic manipulation, the time scale $t_{1}$ for
generating spin squeezing is limited by the adiabatic condition
\end{subequations}
\begin{equation}
\left\vert \frac{\left\langle n\left( t\right) |\dot{d_{n}(t)}\right\rangle
}{E_{n}-E_{d}}\right\vert \ll 1  \label{5-0-2}
\end{equation}%
with eigenstates $\left\vert n\left( t\right) \right\rangle $ (the
corresponding eigen-energy $E_{n}$) and the time-dependent dark states $%
\left\vert d_{n}(t)\right\rangle $ (the corresponding eigen-energy $E_{d}$).
Here,
\begin{equation}
\left\langle n\left( t\right) |\dot{d_{n}(t)}\right\rangle \equiv \frac{%
\left\langle n\left( t\right) \right\vert \frac{\partial }{\partial t}%
H(t)\left\vert d_{n}(t)\right\rangle }{E_{n}-E_{d}}  \label{5-0-3}
\end{equation}%
describes the variation due to the time-dependent Hamiltonian. Thus
the condition for the adiabatic following is $\Omega _{m}\tau \gg
1.$ The typical pulse length $\tau $ is about $150\mu
s$~\cite{Rabi}, which means the
maximum value of the Rabi frequencies $\Omega _{m}$ should much larger than $%
6.67\times 10^{3}Hz$. According to the pulse sequence in Eq. (\ref{5-0-1}),
the time scale $t_{1}$ can be determined when $\theta \left( t_{1}\right) $
is tuned to $\pi /6$ as%
\begin{equation}
t_{1}=\frac{2a+\tau r-\sqrt{4a^{2}+\tau ^{2}r^{2}}}{2r},  \label{5-0-4}
\end{equation}%
where $r=\tanh ^{-1}[{(6-\pi) }/{(6+\pi) }].$When $a$ tends to
infinity, the longest time $t_{1}^{L}\simeq 1/2\tau =75\mu s$ is
obtained.

%
\begin{figure}[tbp]
\begin{centering}
\includegraphics[bb=49 449 568 766,clip,width=3in]{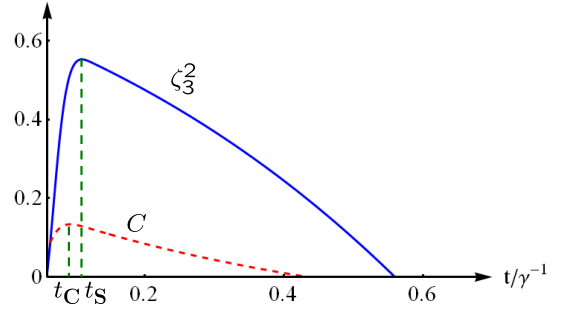}
\par\end{centering}
\caption{(Color online) The evolution of the atomic squeezing $\protect\zeta %
_{3}^{2}$ (blue solid line) and the concurrence $C$ (red dashed
line). The parameters are chosen as $N=20,n=4,K=0$. The maximum of
the atomic squeezing
and the concurrence are respectively obtained at $t_{\mathrm{s}}\approx 0.12\gamma^{-1}=120\mu s$ and $t_{%
\mathrm{c}}\approx 0.09\gamma^{-1}=90\mu s$.} \label{fig:fig14}
\end{figure}


In the other hand, we also assume the exponential decays for the decoherence
processes with the decoherence strength
\begin{equation}
p\left( t\right) =1-e^{-\gamma t},  \label{eq:5-3}
\end{equation}%
which leads to the time scale
\begin{equation}
t_{2}=\frac{1}{\gamma }.  \label{eq:5-4}
\end{equation}%
For example, for the cold trapped $^{87}\mathrm{Rb}$ atomic
ensemble, the typical dephasing damping rate $\gamma =10^{3}Hz$ and
the corresponding dephasing time $t_{2}=1ms$~\cite{Pan,QMR}.

The evolution of the spin-squeezing and the concurrence is shown in Fig.~\ref%
{fig:fig14}. Obviously, The maximum of the atomic squeezing and the
concurrence are respectively obtained at $t_{\mathrm{s}}\approx 0.12\gamma^{-1}=120\mu s$ and $t_{%
\mathrm{c}}\approx 0.09\gamma^{-1}=90\mu s$. Though the
Fig.~\ref{fig:fig14} is drawn under PDC, the optimal times can also
be obtained under ADC and DPC.

\begin{figure}[tbp]
\begin{centering}
\includegraphics[bb=47 454 557 794,clip,width=2.8in]{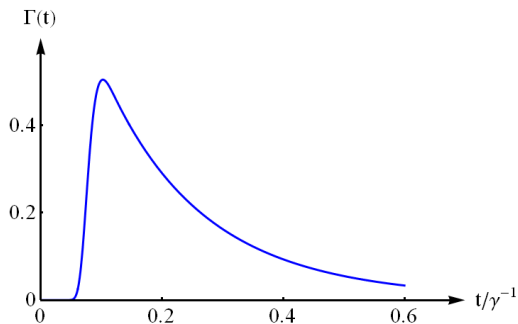}
\par\end{centering}
\caption{(Color online). The retrieval efficiency as a function of
the time. The parameters are chosen as $N=20,n=4,K=0$. The maximum
retrieval efficiency is also obtained around the
$0.12\gamma^{-1}=120\mu s$.} \label{fig:fig15}
\end{figure}

Such competition can also be observed by measuring the cross-correlation~%
\cite{Pan,Pan2}%
\begin{equation}
g_{S,AS}=1+C\Gamma (t)  \label{5-5}
\end{equation}
as a function of the time, where $C$ is a fitting parameter determined by
the excitation probability together with the background noise in the
anti-Stokes channel. Here,

\begin{equation}
\Gamma (t)=\left\vert \left\langle \Psi _{\max }|\Psi (t)\right\rangle
\right\vert ^{2}  \label{5-6}
\end{equation}%
is the retrieval efficiency, which actually characterizes the
complete time-dependence properties of the cross-correlation. For
the dark state we discuss above, to estimate the adiabatical
creation and the decoherence channels at the same time, we choose
$\left\vert \Psi _{\max }\right\rangle $ as the collective atomic
state $|M_{n}\rangle $ and thus the retrieval
efficiency is obtained as%
\begin{equation}
\Gamma (t)=\frac{n!(N-n)!}{N!}\text{ }\frac{_{2}F_{1}(n-N,-n;1;e^{-2\gamma
t})}{\text{ }_{1}F_{1}(-n;N-n+1;-N\cot ^{2}\theta (t))}.  \label{5-7}
\end{equation}%
Here, $_{2}F_{1}(a,b;c;z)$ is generalized Kummer hypergeometric function~%
\cite{Kummer}. The retrieval efficiency is depicted in
Fig.~\ref{fig:fig15}, in which the maximum of the retrieval
efficiency exists when the same condition $t_{1}>t_{2} $ is
satisfied.

%


\section{\label{sec:six}Conclusion}

We investigate the multipartite correlations characterized by the atomic
squeezing beyond the pairwise entanglement in the dark states, which are
proposed for the quantum information storage based on electromagnetically
induced transparency mechanism. The atomic squeezing and the pairwise
entanglement can be created by adiabatically manipulating the Rabi frequency
of the the classical light field on the atomic ensemble. The atomic
squeezing in dark states converts from the sub-Poisson distribution of the
photons. The greatest atomic squeezing and the concurrence can be generated
when the parameter $\theta$ is manipulated to $\pi/2$, which means all the
sub-Poisson distribution of the photons converts to the atomic squeezing.

We also consider the sudden death for the atomic squeezing and the pairwise
entanglement under various decoherence channels. For the three typical
decoherence channels, the sudden deaths of the atomic squeezing happens
later than the sudden death of the concurrence.

According to the above investigation, an optimal time for generating the
greatest atomic squeezing and pairwise entanglement is obtained by studying
in details the competition between the adiabatic creation of quantum
correlation in the atomic ensemble and the decoherences we describe with
decoherence channels.

\begin{acknowledgements} The work is supported by National Natural
Science Foundation of China under Grant Nos. 10547101 and 10604002,
the National Fundamental Research Program of China under Grant No.
2006CB921200. X. Wang is supported by NSFC with grant No.10874151,
10935010, NFRPC with grant No. 2006CB921205; Program for New Century
Excellent Talents in University (NCET), and Science Fundation of
Chinese University.

\end{acknowledgements}


\begin{thebibliography}{99}
\bibitem{Hau} C. Liu, Z. Dutton, C. H. Behroozi, and L. V. Hau, Nature
(London) \textbf{409}, 490 (2001).

\bibitem{Lukin-1} M. D. Lukin, Rev. Mod. Phys. \textbf{75}, 457 (2003).

\bibitem{Lukin-2} M. Fleischhauer and M. D. Lukin, Phys. Rev. A \textbf{65},
022314 (2002).

\bibitem{Lukin-3} M. Fleischhauer, S. F. Yelin, M. D. Lukin, Opt. Commun.
\textbf{179}, 395 (2000).

\bibitem{2003} C. P. Sun, Y. Li, and X. F. Liu, Phys. Rev. Lett. \textbf{91}%
, 147903 (2003).

\bibitem{Dicke} R. H. Dicke, Phys. Rev. \textbf{93}, 99 (1954).

\bibitem{Pan} B. Zhao, Y. A. Chen, X. H. Bao, T. Strassel, C. S. Chuu, X. M.
Jin, J. Schmiedmayer, Z. S. Yuan, S. Chen and J. W. Pan, Nature Phys.
\textbf{5}, 95 (2009).

\bibitem{Kuzmich} Y. O. Dudin, S. D. Jenkins, R. Zhao, D. N. Matsukevich, A.
Kuzmich, and T. A. B. Kennedy, Phys. Rev. Lett. \textbf{103}, 020505 (2009).

\bibitem{APT} M. D. Lukin, S. F. Yelin, and M. Fleischhauer, Phys. Rev.
Lett. \textbf{84}, 4232 (2000).

\bibitem{Noisy} W. H. Zurek, Rev. Mod. Phys. \textbf{75}, 715 (2003).

\bibitem{Random1} C. Mewes and M. Fleischhauer, Phys. Rev. A \textbf{72},
022327 (2005).

\bibitem{Random2}J. Simon and H. Tanji, J. K. Thompson and V. Vuleti\'{c}, Phys. Rev. Lett. \textbf{98},
183601 (2007).

\bibitem{Yuting} T. Yu and J. H. Eberly, Science \textbf{323}, 598 (2009).

\bibitem{Wang-1} X. Wang, A. Miranowicz, Y. X. Liu, C. P. Sun, and F. Nori,
Phys. Rev. A \textbf{80}, 022106 (2010).


\bibitem{Polzik}J. Appel, P. J. Windpassinger, D. Oblak, U. B. Hoff,
N. Kj\ae gaard and E. S. Polzik, Proc. Nat. Acade. U. S. A.
\textbf{106}, 10960 (2009).

\bibitem{Kummer} M. Abramowitz and I. A. Stegun, \textit{Handbook of
Mathematical Functions,} (Dover, New York, 1965).

\bibitem{KU} M. Kitagawa and M. Ueda, Phys. Rev. A \textbf{47}, 5138 (1993).

\bibitem{Wineland} D. J. Wineland, J. J. Bollinger, W. M. Itano, and D. J.
Heinzen, Phys. Rev. A \textbf{50}, 67 (1994).

\bibitem{Toth}G. T\'{o}th, C. Knapp, O. G\"{u}hne, and H. J. Briegel, Phys.
Rev. Lett. \textbf{99}, 250405 (2007); Phys. Rev. A \textbf{79},
042334 (2009).

\bibitem{Sorensen}A. S\o rensen, L. M. Duan, J. I. Cirac, and P. Zoller,
Nature (London) \textbf{409}, 63 (2001).

\bibitem{Wang-2} X. Wang and B. C. Sanders, Phys. Rev. A \textbf{68}, 012101
(2003).

\bibitem{Wootters} W. K. Wootters, Phys. Rev. Lett. \textbf{80}, 2245 (1998).

\bibitem{Wang-3}X. Wang and K. M\o lmer, Eur. Phys. J. D \textbf{18}, 385
(2002).

\bibitem{photonic} C. Santori, D. Fattal, K. C. Fu, P. E. Barclay, and R. G. Beausoleil, New J. Phys. \textbf{11}, 123009 (2009).

\bibitem{decoherence} S. S. Jang, Y. W. Cheong, J. Kim, and H. W. Lee, Phys.
Rev. A \textbf{74}, 062112 (2006).

\bibitem{Rabi} M. Weitz, B. C. Young and S. Chu, Phys. Rev. Lett. \textbf{73}%
, 2563 (1994).

\bibitem{Gaussian} R. G. Unanyan, B. W. Shore, and K. Bergmann, Phys. Rev. A \textbf{63}%
, 043405 (2001).

\bibitem{solitonary} I. R. Sol\'{a}, V. S. Malinovsky, and D. J. Tannor, Phys. Rev. A \textbf{60}%
, 3081 (1999).

\bibitem{QMR} C. Simon, M. Afzelius, J. Appel, A. Boyer de la Giroday, S.J. Dewhurst, N. Gisin, %
C.Y. Hu, F. Jelezko, S. Kroll, J.H. Muller, J. Nunn, E. Polzik, J. Rarity, H. de Riedmatten, %
W. Rosenfeld, A.J. Shields, N. Skold, R.M. Stevenson, R. Thew, I. Walmsley, M. Weber, H. Weinfurter, %
J. Wrachtrup, and R.J. Young, arXiv:1003.1107.

\bibitem{Pan2}Shuai Chen, Yu-Ao Chen, Thorsten Strasse, Zhen-Sheng Yuan, Bo Zhao,
J\"{o}rg Schmiedmayer, and Jian-Wei Pan, Phys. Rev. Lett.
\textbf{97}, 173004 (2006).
\end{thebibliography}
\end{document}